\documentclass[12pt,draftcls,onecolumn]{IEEEtran}
\usepackage{graphicx}
\usepackage{stfloats}
\usepackage{amsmath}
\usepackage{amssymb}
\usepackage{array}
\usepackage[caption=false,font=footnotesize]{subfig}
\usepackage[numbers,sort&compress]{natbib}
\usepackage{url}
\renewcommand{\tablename}{ALGORITHM~}
\newcommand{\tabincell}[2]{\begin{tabular}{@{}#1@{}}#2\end{tabular}}

\begin{document}

\bibliographystyle{IEEEtran}

%
\title{Cognitive Random Stepped Frequency Radar with Sparse Recovery}
\author{Tianyao~Huang,
Yimin~Liu,
Huadong~Meng,
Xiqin~Wang
}
\maketitle

\begin{abstract}
Random stepped frequency (RSF) radar, which transmits random-frequency pulses, can suppress the range ambiguity, improve convert detection, and possess excellent electronic counter-countermeasures (ECCM) ability \cite{Axelsson2007}. In this paper, we apply a sparse recovery method to estimate the range and Doppler of targets. We also propose a cognitive mechanism for RSF radar to further enhance the performance of the sparse recovery method. The carrier frequencies of transmitted pulses are adaptively designed in response to the observed circumstance. We investigate the criterion to design carrier frequencies, and efficient methods are then devised. Simulation results demonstrate that the adaptive frequency-design mechanism significantly improves the performance of target reconstruction in comparison with the non-adaptive mechanism.
\end{abstract}
\begin{IEEEkeywords}
Random stepped frequency radar, adaptive waveform design, sparse recovery, Subspace Pursuit, compressed sensing
\end{IEEEkeywords}

%
\IEEEpeerreviewmaketitle

\section{Introduction}
Since stepped frequency (SF) waveforms can synthesize {very wide} frequency bands with a narrow bandwidth receiver, they are widely used in {{radars}} to generate high-range-resolution {{profiles}} (HRRPs), Synthetic Aperture Radar (SAR) imaging, Inverse SAR (ISAR) imaging, etc. \cite{Wehner1995,Hua1993,Liu2010}. In SF radar, the carrier frequencies of pulse trains are linearly varied with a constant frequency step, which produces a ridge in the range-Doppler ambiguity function \cite{Axelsson2007}. When the transmitted frequencies are changed randomly, rather than linearly, the ridge ambiguity function can be enhanced to a thumbtack function \cite{liu2008}. \par
\begin{figure}[!b]
\centering
\includegraphics[width=3in]{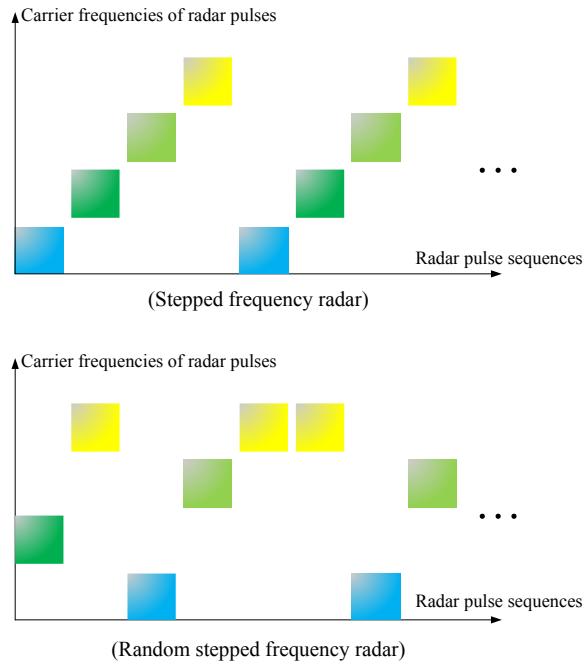}
\caption{Carrier frequencies of stepped frequency (SF) radar and random stepped frequency (RSF) radar.}
\label{fig:rsf}
\end{figure}

In random stepped frequency (RSF) radar, the carrier frequencies are randomly chosen from a given bandwidth \cite{Axelsson2007}; see Fig. \ref{fig:rsf} for the comparison of radar waveforms between SF and RSF radar. Compared with linear SF radar, RSF radar further improves the range-Doppler resolution, suppresses the range ambiguity, and decouples the range and the Doppler \cite{Huang2012}. This technique is attractive for its merits on electronic counter-countermeasures (ECCM) \cite{liu2008} and significantly reduces interference between adjacent radar systems \cite{Axelsson2007}. The RSF waveforms were implemented in a wide-angle SAR to mitigate aliasing artifacts \cite{Luminati2004,Luminati2007} and were used in an ISAR to suppress the Doppler ambiguity \cite{liu2000joe}. In this paper, we focus on estimating the {{ranges}} and Doppler of multiple targets with RSF radar. \par
Sparse recovery and compressed/compressive sensing (CS) have received significant attention in radar signal processing \cite{Gogineni2011,Gurbuz2012,Yu2012,Gogineni2012}. {{By exploiting the sparsity, the theory of CS promises to exactly recover a sparse vector of length $N$ with high probability from much fewer than $N$ measurements \cite{Yu2012}.
In applications of RSF radars, the number of targets in the same coarse range bin is usually small, which forms a {\it{sparse}} scenario. The}} CS methods are applicable to detect the targets and recover the {{ranges}}, velocities and scattering intensities of the targets.
\par
\begin{figure}[!h]
\centering
\includegraphics[width=3in]{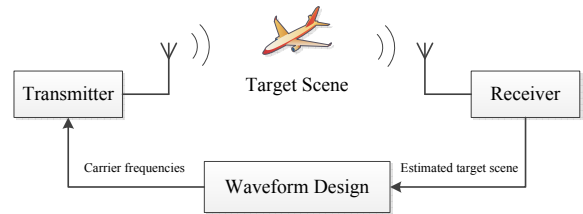}
\caption{The framework of cognitive RSF radar \cite{Zhang2012}.}
\label{fig:cogrsf}
\end{figure}

{{In order to enhance the performance with sparse modeling, the idea of cognitive radar is introduced to make use of the priori information of the target scenario. }}
Cognitive radar was first proposed in \cite{Haykin2006} and has attracted increasing research interest for a number of years \cite{Conte2006,Inggs2010,Wei2011}. In these literatures, cognitive radar is defined as a radar that can adaptively vary the transmission waveform according to the environmental information obtained by the radar. By exploiting the circumstance information, cognitive radar provides a significant improvement on radar performance. In the context of RSF radar, we adaptively design frequencies of transmitted pulses according to the observed target scene to further improve the performance of compressed sensing and pursue more accurate reconstruction of targets. The framework of cognitive RSF radar is demonstrated in Fig. \ref{fig:cogrsf}. \par
Similar ideas, applying the cognitive concept to compressed sensing radar, are also found in \cite{Sen2011,Zhang2012,Gogineni2011}, which significantly improve in performance over the corresponding radars without cognitive mechanism. Sen et al. \cite{Sen2011} adaptively design the amplitudes of the transmitted subcarriers in an orthogonal-frequency-division multiplexing (OFDM) radar. Gogineni et al. \cite{Gogineni2011} develop an adaptive energy-allocation mechanism for different transmitting antennas of multiple-input multiple-output (MIMO) radar system. Zhang et al. \cite{Zhang2012} optimize the sensing matrix and the phases of continuous phase-coded waveforms. However in this paper, we target optimizing the carrier frequencies of transmitted signals in RSF radar{. The optimization parameters differ from those} in \cite{Sen2011,Zhang2012,Gogineni2011}, and these waveform-design methods are not directly applicable in RSF radar.
{ Gogineni et al. \cite{Gogineni2012a} and Han et al. \cite{Han2013} consider the carrier frequencies design problem with a sparse model for a frequency-hopping MIMO radar. In \cite{Gogineni2012a,Han2013}, the carrier frequencies are designed to reduce the block coherence measure of the sensing matrix. The block coherence is regardless of the target scenario. While in our paper, the carrier frequencies are adaptively optimized to fit the target scenario. The priori information of the targets is exploited to improve the reconstruction performance.}
\par
Aiming at reducing the reconstruction errors of RSF radar, the criterion, minimizing the Cramer-Rao bound (CRB) (or Cramer-Rao lower bound, CRLB) of sparse recovery is applied to design the carrier frequencies.
The performance of RSF radar can be affected by the target scenario if the target returns interfere each other. The CRB depends on the target scenario and the carrier frequencies. The CRB can be seen as a measure of the interference as shown later in Subsection \ref{subsec:obj}.  Minimizing the CRB by designing the carrier frequencies reduces the interferences between target returns and thus enhances the recovery performance.
Considering different potential applications of the cognitive scheme, we devise several efficient algorithms to calculate the optimal transmitting frequencies of radar pulses.
For computational convenience, an approximation to the CRB criterion is also proposed for RSF radar. The consistency between two criterions is analyzed.
\par
The rest of the paper is organized as follows. In Section \ref{Sec:signalmodel}, the echo signal model of RSF radar is introduced. In Section \ref{Sec:SR}, we apply a compressed sensing algorithm to reconstruct the target scene and introduce some research results on the lower bound of sparse recovery errors. Then, in Section \ref{Sec:cog}, we present an adaptive waveform design approach to reduce the lower bound. The merits of the proposed mechanism are demonstrated in Section \ref{Sec:sim} with some simulation results. Section \ref{Sec:conclusion} is devoted to a brief conclusion.\par
\section{Radar Echo Signal Model}{\label{Sec:signalmodel}}
In this section, we describe the signal model of RSF radar. In each coherent processing interval (CPI), $N$ monotone pulses are transmitted with a constant pulse repetition interval $T$.
{he duration of each pulse is $T_p$.
The synthetic bandwidth of the baseband is $B$. The frequency of the $n$th pulse is $f_n \in\left[f_{c}, f_{c} + B\right]$, $n = 0,1 \dots, N-1$, where $f_c$ is the central carrier frequency.
Random frequency is set as $f_n = f_{c} + d_{n} \Delta f$, where $\Delta f$ is the frequency step size, and $d_n$ is a random integer between 0 and the floor integer $\lfloor B/\Delta f \rfloor$.
To avoid 'ghost image' phenomenon, the frequency step size $\Delta f$ should be less than $1/T_p$ \cite{Liu2009}. This is further discussed in the next-to-last paragraph of this section. Actually, the frequency step size can be rather small benefiting from the development of Direct Digital Synthesizer (DDS) technique. For example, for a bandwidth $B = 400$ MHz, a frequency resolution of $\Delta f \approx 0.23$ Hz can be achieved using the AD9910 \cite{AD9910}. In this case, the integer $\lfloor B/\Delta f \rfloor$ is huge.
For notational brevity, the carrier frequency is rewritten as $f_n = f_{c} + c_{n} B$, where $c_n = d_n\Delta f/B \in [0,1]$ is called as the $n$th {\it{frequency-modulation code}}. Since $B/\Delta f $ is huge, $c_n$ is assumed as a continuous, real number.
The $n$th transmitted pulse is described as
\begin{equation}{\label{Equ:tx}}
T_{x}(n,t)={\text{rect}}\left(\frac {t-nT}{T_{p}}\right)e^{j2\pi \left( f_{c} + c_{n}{{B}} \right) \left( t-nT \right) },
\end{equation}
where {rect$(\cdot)$ is a rectangular function defined as}
\begin{equation}
{\text{rect}}(x)=\left\{
\begin{array}{l}
1,0 \leq x \leq 1,\\
0,{\text{otherwise.}}
\end{array}
\right.
\end{equation}
\par
Based on the "stop and hop" assumption, the echo of the $n$th pulse from {a scatterer} is
\begin{equation}{\label{Equ:rxnt}}
{R_x}\left( {n,t} \right) \approx  {{{{\beta}}}T_{x}\left(n,t-\frac{2{r\left(t\right)}}{c} \right)},
\end{equation}
where $c$ is the wave propagation speed. $\beta$ and $r(t)$ are the scattering intensity and the range of the target at instant $t$ with respect to the radar, respectively.
Suppose the target is moving radially at a constant speed $v$, then $r(t) = r(0)+vt$.
In this paper, it is simply assumed that tangential or rotational motion, and acceleration (and higher-order terms) of the target are ignorable or have been compensated previously. Refer to \cite{Huang2012} and the references therein for details of motion compensation for RSF radar.
\par
{The $n$th echo is sampled at fast-time instant $t_{s}(n,l_r) = nT + l_r/f_s$, where $f_s$ is the sampling rate and $l_r = 1,2,\dots,L_r$ ($L_r<T f_s$) denotes the index of samples. In the case that monotone pulses are transmitted, the sampling rate should be no less than $1/T_p$ such that no return will be missed. It is set as $f_s = 1/T_p$ in this paper. At $t_{s}(n,l_r)$, echoes from targets located between $r_s(l_r-1)$ and $r_s(l_r)$ will be sampled, where $r_s(l_r) = l_r/f_s \cdot c/2$ denotes the range corresponding to the $l_r$th sampling time instant; see Fig. \ref{fig:sample}. The zone $[r_s(l_r-1),r_s(l_r)]$ is called as a coarse-range bin.} \par
\begin{figure}[!h]
\centering
\includegraphics[width=3in]{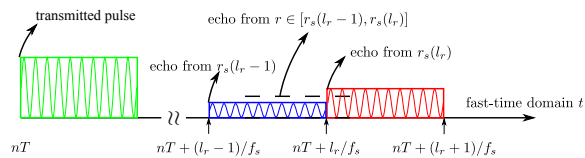}
{\caption{Fast-time domain sampling. The sampling period is set as the pulse width $1/f_s = T_p$. As shown in the figure, after transmission of the radar pulse, the echo from the target located at $r_s(l_r-1)$ arrives previously to echoes from $r>r_s(l_r-1)$. Echo from targets located inside $[r_s(l_r-1),r_s(l_r)]$ are sampled at $t_s(n,l_r) = nT+l_r/f_s$. }
\label{fig:sample}}
\end{figure}
\par
\begin{figure}[!h]
\centering
\includegraphics[width=3in]{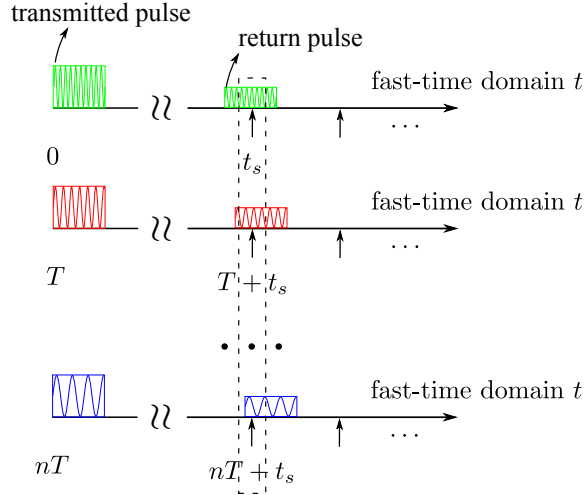}
{\caption{Samples at fast-time instant $nT+t_s$ comprise a serial of measurements, $n = 0,1,\dots,N-1$. }
\label{fig:cpi}}
\end{figure}
Samples of successively transmitted pulses from the same coarse-range bin are collected; see Fig. \ref{fig:cpi}. These data form the measurements within a CPI, and are used to generate HRRPs of the coarse-range bin. Samples from different coarse-range bins are processed individually. This paper focuses on HRRP, and without loss of generality only one coarse-range bin is considered. For notational brevity, the index $l_r$ of the coarse-range bin is omitted in the rest of paper. For example, $t_s(n,l_r)$ is simplified as $t_s(n)$.
Substitute $t = t_s(n)$ and $r_s = (t_s(n) - nT)c/2$ into (\ref{Equ:tx}) and (\ref{Equ:rxnt}). The sampled $n$th echo (\ref{Equ:rxnt}) can be expressed as
\begin{equation}{\label{Equ:echo1}}
\begin{split}
{R_x}\left( {n} \right) &=  {{\beta}e^ {  j2\pi \left(f_c+c_nB\right)\left( t_s - \frac{2r(t_s)}{c} -nT \right)}}\\
&{={{\beta}e^ {  j4\pi \left(f_c+c_nB\right)\frac{r_s - r(t_s)}{c} }}}\\
&{={{\beta}e^ {  -j4\pi \left(f_c+c_nB\right)\frac{r(0) -r_s + vnT + 2vr_s/c}{c} }}}.
\end{split}
\end{equation}
Since only one coarse-range bin is considered, the fast-time parameter $t_s$ is omitted in (\ref{Equ:echo1}) and the rest of paper for simplicity. $R_x(n,t_s)$ is replaced by $R_x(n)$. Denote $R = r(0)-r_s$ as the high-resolution range of the target and ignore the term $2vr_s/{c^2}$, then
\begin{equation}{\label{Equ:echo2}}
\begin{split}
{R_x}\left( {n} \right) &\approx {{\beta}e^ {  -j4\pi \left(f_c+c_nB\right)\left( R+vnT \right)/c }}\\
&{= {{\beta} e^ {  - j\frac{4\pi f_cR}{c} - j{ {\frac{4\pi B R}{c}}c_n} - j\frac{{{4\pi {f_c}vT}}}{c}nc'_{n}}}},
\end{split}
\end{equation}
where $c'_{n} =  1 + c_{n} {{{B}}}/f_c $.
\par
{When there are $K$ targets inside a coarse-range bin, the received signals are recast as linear combinations of echoes from different targets and (\ref{Equ:echo2}) is rewritten as}
\begin{equation}{\label{Equ:echo}}
\begin{split}
{R_x}\left( {n} \right) \approx \sum\limits_{k=1}^{K} {{\gamma _k}\exp { \left( jp_kc_n + jq_knc'_{n}\right)}},
\end{split}
\end{equation}
{where ${\gamma _k} = {\beta _k}\exp \left( { - j{4\pi {f_c}}}{R_k}/c \right)$, ${p_k} =  - {4\pi B R_k}/c$ and ${q_k} =  - {{4\pi {f_c}v_kT}}/c$ are used for notational brevity. $\beta_k$, $R_k$ and $v_k$ are the scattering magnitude, high-resolution range and velocity of the $k$th target, respectively. Since $|\gamma_k|$, $p_k$ and $q_k$ are proportional with $|\beta_k|$, $R_k$ and $v_k$, respectively, in the remainder of this paper they are simply regarded as the scattering coefficient, range and Doppler parameter of the $k$th target, respectively.}
\par
{The HRRPs are synthesized with the observations in the form of (\ref{Equ:echo}).
Substitute $c_n = d_n\Delta f/B$ into (\ref{Equ:echo}), it can be implied that the unambiguous scope of HRRPs is $c/(2\Delta f)$ \cite{Axelsson2007}. It should be larger than the scope of a coarse-range bin $cT_p/2$ in order to avoid ghost image \cite{Liu2009}, which yields $\Delta f<1/T_p$ as stated in the start of this section.
After} we obtain all of the sampled data at the same fast-time instant $t_s$ in a CPI, {these data are then used to reconstruct $\gamma$, $p$ and $q$} of all the targets.
\par
{In many practical cases, radar signal returns are corrupted by thermal noises and clutters. Noise is discussed in ensuing sections, but we simply assume in this paper that the returns have been filtered for clutter reduction prior to synthesizing HRRPs. Refer to \cite{Axelsson2006} for details of clutter cancelation algorithms for RSF radar.
In addition, we limit the scope of this paper to 1-dimensional range profiling only. 2-D imaging (including azimuth dimension) with RSF radar remains for future work. Those readers interested in 2-D imaging with RSF radar are referred to \cite{Huang2012} and references therein.}

\section{Sparse Recovery}{\label{Sec:SR}}

\subsection{Sparse Modeling}
We discretize the possible range and Doppler values of the targets, i.e., $p$ and $q$ in (\ref{Equ:echo}), into $P$ and $Q$ grid points, respectively. Thus, we have $PQ$ possible range-Doppler pairs $(p_l,q_l)$, $l = 1,2,\dots,PQ$. Then, we can rewrite (\ref{Equ:echo}), the signal that the radar receives, as a combination of echoes from all possible targets,
\begin{equation}{\label{Equ:echogrid1}}
{R_x}(n) = \sum\limits_{l=1}^{PQ} {{\gamma _l} \exp { \left( jp_lc_n + jq_lnc'_{n}\right)}},
\end{equation}
where $\gamma_l$ denotes the scattering coefficient of the target presented at $(p_l,q_l)$. If no target exists at $(p_l,q_l)$, $\gamma _l =0$. We define a vector ${\boldsymbol \phi}_l \in \mathbb{C}^N$, in which the $n$th element is
\begin{equation}{\label{Equ:echogrid}}
{\phi}_l(n) = \exp { \left( jp_lc_n + jq_lnc'_{n}\right)}.
\end{equation}
Note that the element ${\boldsymbol \phi}_l(n)$ is built from the modulation code $c_n$. Form a matrix ${\boldsymbol \Phi} = \left[ {\boldsymbol \phi}_1,{\boldsymbol \phi}_2,\dots,{\boldsymbol \phi}_{PQ} \right] \in \mathbb{C}^{N \times PQ}$ and modulation code sequence ${\bf c} = \left[c_0,c_1,\dots,c_{N-1} \right]^{\rm T} \in \mathbb{R}^{N}$, where $(\cdot)^{\rm T}$ denotes the transpose of a matrix or a vector. Note that ${\boldsymbol \Phi}({{\bf c}})$ depends on $\bf c$. Unless specifically stated in the rest of paper, we use ${\boldsymbol \Phi}$ instead of ${\boldsymbol \Phi}({{\bf c}})$ for simplicity. Generate ${\bf x} = \left[\gamma_1,\gamma_2 ,\dots,\gamma_{PQ} \right]^{\rm T} \in \mathbb{C}^{PQ}$, which represents the scattering coefficients. Assume the echoes are corrupted by additive Gaussian white noise; thus, the received echoes ({{\ref{Equ:echogrid1}}}) can be written in a matrix form as
\begin{equation}{\label{equ:matrix}}
{\bf y} = {\boldsymbol \Phi}{\bf x} + {\bf w},
\end{equation}
where ${\bf y} \in \mathbb{C}^N$ represents the corrupted echoes and ${\bf w}$ is a noise vector with a complex normal distribution $\mathbb{C}\mathcal{N} \left( {\bf 0},\sigma^2{\bf I}_N \right)$. $\sigma^2$ denotes the variance of the noise and ${\bf I}_N$ denotes an identity matrix with dimension of $N$. ${\boldsymbol \Phi}$ is often referred to as a {\it dictionary matrix} in literatures on compressed sensing.
${\bf x}$ is an unknown vector to be recovered. ${\bf x}$ is $K$-sparse, which means that there are only $K$ nonzero or $K$ prominent elements in ${\bf x}$. Given $\bf y$ and $\boldsymbol \Phi$, sparse recovery solves $\bf x$ with constraints on the sparseness of $\bf x$. When $\bf x$ is recovered, the high-resolution ranges and Doppler of the targets can be inferred from the support set $\Lambda = {\rm supp}({\bf x})$, where ${\rm supp}(\cdot)$ denotes the set that consists of the indices of nonzero elements in the vector.

\subsection{Sparse Recovery}
Sparse recovery algorithms estimate ${\bf x}$ in (\ref{equ:matrix}) by exploiting its sparsity.
In this paper, {{Subspace Pursuit (SP) \cite{Dai2009} is adopted as the sparse recovery algorithm. The SP algorithm is a kind of greedy approach \cite{Dai2009,Huang2012a}, and possesses a provable reconstruction capability comparable to that of Basis Pursuit \cite{chen2001atomic} and the Dantzig Selector \cite{candes2007dantzig} approaches and a low computational complexity similar to that of the Orthogonal Matching Pursuit (OMP) \cite{Davenport2010} and Regularized Orthogonal Matching Pursuit (ROMP) \cite{Needell2009} approaches.}}
More precisely, SP is a greedy approach to solve the $\ell _0$ minimization problem
\begin{equation}{\label{minl0}}
\min \| {\bf x} \|_0, {\rm \ subject \ to \ } \| {\bf y}- {\boldsymbol \Phi}{\bf x}\|_2<\eta,
\end{equation}
where $\| \cdot \|_0$ and $\| \cdot \|_2$ denote the $\ell _0$ and Euclidean ($\ell _2$) norm of a vector, respectively. $\eta$ represents the power of noise.\par
We recall the main steps of the SP algorithm \cite{Dai2009} in Algorithm \ref{Alg:SP}, where ${\bf x}_{\Lambda}/{\boldsymbol \Phi}_{\Lambda}$ denotes a sub-vector/matrix that consists of entries/columns indexed in the set $\Lambda$. $(\cdot)^{\dagger}$ denotes the Moore-Penrose pseudo inverse, i.e., ${\bf A}^{\dagger} = ({\bf A}^{\rm H}{\bf A})^{-1}{\bf A}^{\rm H}$, where $(\cdot)^{\rm H}$ is the Hermitian transpose.
    \begin{table}[h]{\caption{The Standard SP Algorithm}\label{Alg:SP}}\centering
     \begin{tabular}{l}\hline
              \tabincell{l}{ 1) Input $K,{\bf y},{\bf \Phi}$. Set the support set ${\Lambda ^{\left( 0 \right)}} = \emptyset $, the residual error ${{\bf{r}}^{\left( 0 \right)}} $\\$= {\bf{y}}$, and the iteration counter $i=0$.}  \\
              \tabincell{l}{2) Calculate the correlations ${\bf p} =  {\boldsymbol \Phi}^{\rm H} {\bf r}^{(i)}$.} \\
              3) Merge the set ${\tilde{\Lambda}} = {\Lambda^{(i)}} \cup \{K$ indices corresponding to the largest \\ magnitude entries in ${\bf p} \}$. \\
              4) Set ${\bf x}_p = {\boldsymbol \Phi}_{\tilde{\Lambda}} ^{\dagger}{\bf y}$, and update the set $\Lambda^{(i+1)} = \{K$ indices \\ corresponding to the largest entries in ${\bf x}_p \}$.\\
              5) Update the residual error ${\bf{r}}^{(i+1)} = {\bf{y}} - {\boldsymbol{\Phi}}_{\Lambda^{(i+1)}}{\boldsymbol{\Phi}}_{\Lambda^{(i+1)}}^{\dagger}{\bf{y}}$.\\
              6) Increase $i$. Return to Step 2 until stop criterion, e.g., $\Lambda^{(i)} = \Lambda^{(i-1)}$, \\ is satisfied.\\
              \tabincell{l}{7) Output ${\hat{\bf{x}}}$, where ${\hat{\bf{x}}}_{\Lambda^{(i)}} = {\boldsymbol{\Phi}}_{\Lambda^{(i)}}^{\dagger}{\bf{y}}$ and the rest entries are all zeros.}\\ \hline
          \end{tabular}
    \end{table}
\par
We cite here some brief lower bound analysis on sparse recovery errors, and in Section \ref{Sec:cog} we target adaptively reducing the lower bound via designing radar waveforms. If the lower bound can be achieved by some sparse recovery methods, reduction in the bound yields decrease in errors of these recovery methods. Denote a solution to the sparse recovery problem in (\ref{minl0}) as ${\hat {\bf x}}$; thus, the recovery error can be described as $\| {\bf x} -{\hat{\bf x}} \|_2^2$.
The Cramer-Rao bound (CRB) or Cramer-Rao lower bound (CRLB) is a well-known tool in estimation theory that expresses a lower bound on the variance of any unbiased estimator of an unknown deterministic parameter \cite{Kay1993}.
For estimating the sparse vector $\bf x$ of $\|{\bf x}\|_0 = K$,  Ben-Haim etc. \cite{Ben-Haim2010} derive the constrained CRB as
\begin{equation}{\label{Equ:CRBMSE}}
{\rm E} \left[ \| {\bf x} - {\hat {\bf x}} \|_2^2 \right] \geq \sigma^2 {\rm tr} \left( ({\boldsymbol \Phi}_{\Lambda ^*}^{\rm H}{\boldsymbol \Phi}_{\Lambda ^*})^{-1} \right), \ \|{\bf x}\|_0=K,
\end{equation}
where E$[\cdot]$ denotes the expectation of a random variable, tr$(\cdot)$ denotes the trace of a matrix and $\sigma^2$ is the variance of noise in (\ref{equ:matrix}). $\Lambda ^{*}$ is the true support set. Ben-Haim etc. also state that the constrained CRB can be attained when a large number of independent measurements are available via the Maximum-Likelihood approach \cite{Ben-Haim2010}
\begin{equation}{\label{Equ:benml}}
{\hat{\bf x}} = \mathop {\arg \min} \limits_{\bf x} \| {\bf y} - {\boldsymbol \Phi}{\bf x} \|_2^2,\text{ subject to }\|{\bf x}\|_0 \leq K.
\end{equation}
In \cite{candes2007dantzig}, Candes presents an oracle sparse estimator, in which a Genie provides the true support set $\Lambda^*$. Denote ${\hat {\bf x}}_{\rm oracle}$ as the oracle estimate of $\bf x$. The mean squared error (MSE) on the estimation of $\bf x$ is
\begin{equation}{\label{Equ:oraclemse}}
{\rm E} \left[ \| {\bf x} - {\hat {\bf x}}_{\rm oracle} \|_2^2 \right] = \sigma^2 {\rm tr} \left( ({\boldsymbol \Phi}_{\Lambda ^*}^{\rm H}{\boldsymbol \Phi}_{\Lambda ^*})^{-1} \right).
\end{equation}
For any unbiased estimator $\hat{\bf x}$ of $\bf x$, the variance of the error ${\rm E} \left[ \| {\bf x} - {\hat {\bf x}} \|_2^2 \right] \geq {\rm E} \left[ \| {\bf x} - {\hat {\bf x}}_{\rm oracle} \|_2^2 \right]$ \cite{Babadi2009}, which is the same as the constrained CRB in (\ref{Equ:CRBMSE}).\par
The achievability of the CRB for noisy sparse recovery has been reported in \cite{candes2007dantzig,Babadi2009,Niazadeh2012}. The Dantzig Selector \cite{candes2007dantzig}, which is based on linear programming, achieves the error in (\ref{Equ:oraclemse}) up to a factor of log$(PQ)$. Note that in the Dantzig Selector, a priori knowledge of $K$ is not necessary. Babadi et al. \cite{Babadi2009} establish a joint typicality estimator that asymptotically achieves the CRB without any information about the support set $\Lambda^*$ as the number of measurements $N \rightarrow \infty$ for $\boldsymbol \Phi$, a random Gaussian matrix of which elements are drawn i.i.d. (independent identically distributed) from $\mathcal{N}(0,1)$. Niazadeh et al. \cite{Niazadeh2012} generalize the conditions for the problem of the asymptotic achievability of CRB. They relax the Gaussianity constraint on $\boldsymbol \Phi$ assuming that $\boldsymbol \Phi$ is randomly generated according to a distribution that satisfies some sort of concentration of measures inequality \cite{Niazadeh2012}.
\section{Optimal Code Design}{\label{Sec:cog}}
\subsection{Optimization Criterion}
In cognitive RSF radar, we use a priori information about the target scene to adaptively design the modulation code sequence $\bf c$ to better recover the targets.
{The metric for the recovery performance is mean square errors, ${\rm E} \left[ \| {\bf x} - {\hat {\bf x}} \|_2^2 \right]$, which is commonly used in radar application. MSE is a good way to capture the systems performance. The estimation error of every target is equally indicated in MSE and the MSE is small only when all targets are accurately estimated.
To reduce the MSE, we choose the strategy minimizing the lower bound on the recovery error ${\rm E} \left[ \| {\bf x} - {\hat {\bf x}} \|_2^2 \right]$ in (\ref{Equ:CRBMSE})}
\begin{equation}{\label{Equ:optcode}}
{\bf c}_{\rm opt} = \mathop {\arg \min} \limits_{\bf c} {\rm tr} \left( ({\boldsymbol \Phi}_{\Lambda}^{\rm H}{\boldsymbol \Phi}_{\Lambda})^{-1} \right) = \mathop {\arg \min} \limits_{\bf c} {\rm tr} \left( ({\bf A}^{\rm H}{\bf A})^{-1} \right),
\end{equation}
where ${\bf A} = \frac{1}{\sqrt{N}}{\boldsymbol \Phi}_{\Lambda} \in \mathbb{C}^{N \times |\Lambda|}$, and $|\Lambda|$ denotes the cardinality of $\Lambda$.
Note that the sub-dictionary ${\boldsymbol \Phi}_{\Lambda}$ and $\bf A$ are functions of code sequence ${\bf c}$; see (\ref{Equ:echogrid}). Since the correct support set $\Lambda ^*$ is actually unknown, we use the {previous} estimate $\Lambda = {\rm supp}({\hat {\bf x}})$ instead.
\par
{From the perspective of the estimation theory,} CRB is the inverse of the Fisher information \cite{Kay1993}, which can be seen as a measure of efficiency of the sensing system ${\bf y} = {\boldsymbol \Phi}{\bf x}+{\bf w}$. The sensing system can be more informative if ${\boldsymbol \Phi}$ is designed to lower the CRB.
{From the perspective of radar system, the carrier frequencies are designed to adapt to the target scenario. Minimizing the CRB reduces interference between target returns and contributes to the reconstruction of targets. This is further discussed in the next-to-last paragraph in Subsection \ref{subsec:obj}.
Simulation results in Subsection V-A demonstrate the effect of the CRB criterion on enhancing the reconstruction performance.}
\par
In Subsection \ref{subsec:obj}, we propose an approximation of the objective function in (\ref{Equ:optcode}) for computability. For different potential applications of the cognitive mechanism, we develop two types of efficient algorithms to calculate the optimal modulation codes in Subsection \ref{subsec:steepest} and \ref{subsec:sequential}, respectively.
In Subsection \ref{subsec:steepest}, we develop a steepest descent method for batch-oriented codes optimization, in which a batch of codes are designed in each optimization operation and the previous measurements are not used again. Subsection \ref{subsec:sequential} describes a sequential code design method that designs only one code in one operation; the previous measurements are reused. The batch-oriented and sequential operation modes are demonstrated in Fig. \ref{fig:batch} and Fig. \ref{fig:sequential}, respectively.\par
\begin{figure}[!h]
\centering
\includegraphics[width=3in]{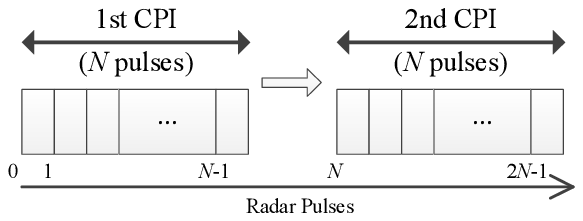}
\caption{Batch-oriented code design. In each code-optimization operation, a batch of codes, which constitutes a CPI, are designed based on the estimation results with last CPI data. The previous measurements are not used again.}
\label{fig:batch}
\end{figure}
\begin{figure}[!h]
\centering
\includegraphics[width=1.7 in]{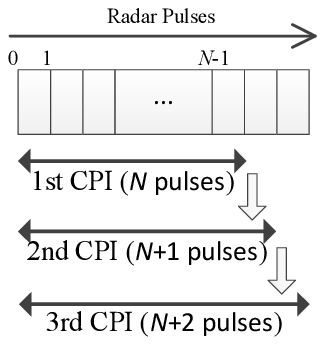}
\caption{Sequential code design. In each code-optimization operation, only one code is designed based on the estimation results with last CPI data. The latest radar pulse and last CPI data constitute a new CPI, which means that measurements are reused.}
\label{fig:sequential}
\end{figure}
\subsection{Objective Function Approximation}\label{subsec:obj}
To apply steepest descent method to solve the optimal problem (\ref{Equ:optcode}), we need to calculate the gradient vector of ${\rm tr} \left( ({\bf A}^{\rm H}{\bf A})^{-1} \right)$ with respect to $\bf c$. {This} task is quite difficult because the matrix inverse is involved. In this subsection, the original objective function in (\ref{Equ:optcode}) is approximated by
\begin{equation}{\label{Equ:optcode2}}
{\bf c}_{\rm opt} = \mathop {\arg \min} \limits_{\bf c} {\rm tr} \left( {\bf A}^{\rm H}{\bf A}{\bf A}^{\rm H}{\bf A} \right),
\end{equation}
such that it is easier to calculate the gradient vector because of elimination of the matrix inverse operation.
\par
${\rm tr} \left( {\bf A}^{\rm H}{\bf A}{\bf A}^{\rm H}{\bf A} \right)$ is adopted because it has a performance similar to ${\rm tr} \left( ({\bf A}^{\rm H}{\bf A})^{-1} \right)$ with respect to $\bf c$ in the scenario of RSF radar. Actually, both objective functions aim at forcing the eigenvalues of ${\bf A}^{\rm H}{\bf A}$ to approach 1, which is explained in the rest of this subsection. \par
In the original optimal problem (\ref{Equ:optcode}),
\begin{equation}{\label{Equ:eigA}}
{\rm tr} \left( ( {\bf A}^{\rm H}{\bf A})^{-1} \right)
= \sum \limits_{i=1}^{\mid \Lambda \mid}{\frac{1}{\lambda _i}} \geq \frac{{\mid \Lambda \mid}^{2}}{\sum \limits_{i=1}^{\mid \Lambda \mid}{\lambda _i}},
\end{equation}
where $\lambda _i$ is the $i$th eigenvalue of $ {\bf A}^{\rm H}{\bf A}$, i.e., ${\bf A}^{\rm H}{\bf A} = {\bf U}{\boldsymbol \Sigma}{\bf U}^{\rm H}$. ${\bf U} \in \mathbb{C}^{|\Lambda| \times |\Lambda|}$ is the characteristic matrix of ${\bf A}^{\rm H}{\bf A}$, which is a unitary matrix ${\bf U}{\bf U}^{\rm H} = {\bf U}^{\rm H}{\bf U} = {\bf I}_{|\Lambda|}$. ${\boldsymbol \Sigma} = {\rm diag}\left({\left[\lambda _1,\lambda _2,\dots,\lambda _{|\Lambda|}\right]}\right) \in \mathbb{R}^{|\Lambda| \times |\Lambda|}$, where diag$(\cdot)$ denotes a diagonal matrix with diagonal elements as indicated in the given vector.
The inequality in (\ref{Equ:eigA}) holds because the harmonic mean ${|\Lambda|}/{\sum \limits_{i=1}^{\mid \Lambda \mid}{\frac{1}{\lambda _i}}}$ is less than or equal to the arithmetic mean ${\sum \limits_{i=1}^{\mid \Lambda \mid}{{\lambda _i}}}/{|\Lambda|}$. Two means are equal if and only if there is a code sequence $\bf c$ such that
\begin{equation}
\lambda_1 = \lambda_2 = \cdots = \lambda_{|\Lambda|}.
\end{equation}
Referring to (\ref{Equ:echogrid}) and (\ref{Equ:optcode}), since the $l$th diagonal element of ${\bf A}^{\rm H}{\bf A}$ equals 1, i.e.,
\begin{equation}{\label{Equ:diagelement}}
\begin{split}
\left( {\bf A}^{\rm H}{\bf A} \right)_{ll} = \sum \limits_{n=0}^{N-1}& \frac{1}{\sqrt{N}}\exp { \left( -jp_lc_n - jq_lnc'_{n}\right)}\\
&\cdot\frac{1}{\sqrt{N}}\exp { \left( jp_lc_n + jq_lnc'_{n}\right)}=1,
\end{split}
\end{equation}
it always holds in the scenario of RSF radar that
\begin{equation}
\sum \limits_{i=1}^{\mid \Lambda \mid}{\lambda _i} = {\rm tr} \left( {\bf A}^{\rm H}{\bf A} \right) = \sum \limits_{l=1}^{\mid \Lambda \mid} {\left({\bf A}^{\rm H}{\bf A}\right)_{ll}} = {\mid \Lambda \mid};
\end{equation}
thus, the equality in (\ref{Equ:eigA}) holds if and only if
\begin{equation}
\lambda_1 = \lambda_2 = \cdots = \lambda_{|\Lambda|} =\frac{\sum \limits_{i=1}^{\mid \Lambda \mid}{\lambda _i}}{|\Lambda|}= 1,
\end{equation}
i.e., ${\boldsymbol \Sigma} = {\bf I}_{|\Lambda|}$. As a result, $\bf A$ is a semi-unitary matrix, i.e.,
\begin{equation}
{\bf A}^{\rm H}{\bf A} = {\bf U}{\boldsymbol \Sigma}{\bf U}^{\rm H} = {\bf U}{\bf U}^{\rm H} = {\bf I}_{|\Lambda|}.
\end{equation}
\par
As discussed above, minimizing the original objective ${\rm tr} \left( ( {\bf A}^{\rm H}{\bf A})^{-1} \right) = \sum \limits_{i=1}^{\mid \Lambda \mid}{\frac{1}{\lambda _i}}$ provides a code sequence $\bf c$ that leads to ${\boldsymbol \Sigma} = {\bf I}_{|\Lambda|}$ if it exists.
When there is no such code sequence that results in a semi-unitary $\bf A$, $\sum \limits_{i=1}^{\mid \Lambda \mid}{\frac{1}{\lambda _i}}$ approaches the minimum if all of the eigenvalues $\lambda_i$ are close to 1 because the eigenvalues obey the constraint $\sum \limits_{i=1}^{\mid \Lambda \mid}{\lambda _i} = {\mid \Lambda \mid}$. We see that the original optimization seeks all eigenvalues close to 1.\par
Note that it is difficult to calculate the gradient of the objective function, minimizing $\sum \limits_{i=1}^{\mid \Lambda \mid}{\frac{1}{\lambda _i}}$ or ${\rm tr} \left( ({\bf A}^{\rm H}{\bf A})^{-1} \right)$, because of the matrix inverse involved. Since the original objective in (\ref{Equ:optcode}) substantially aims at that all eigenvalues approach 1, we can approximately apply least squares to force the eigenvalues close to 1; thus, the initial objective function is replaced with
\begin{equation}{\label{Equ:optcodeLS}}
\begin{split}
{\bf c}_{\rm opt} &= \mathop {\arg \min} \limits_{\bf c} \sum \limits_{i=1}^{\mid \Lambda \mid}{(\lambda _i-1)^2}\\
&= \mathop {\arg \min} \limits_{\bf c} \sum \limits_{i=1}^{\mid \Lambda \mid}{ \lambda_i^2}-\sum \limits_{i=1}^{\mid \Lambda \mid}{\left( 2\lambda_i-1 \right)}\\
&= \mathop {\arg \min} \limits_{\bf c} \sum \limits_{i=1}^{\mid \Lambda \mid}{\lambda_i^2}-|\Lambda|\\
&= \mathop {\arg \min} \limits_{\bf c} {\rm tr} \left( {\bf A}^{\rm H}{\bf A}{\bf A}^{\rm H}{\bf A} \right),
\end{split}
\end{equation}
where $\sum \limits_{i=1}^{\mid \Lambda \mid}{\lambda _i} = {\mid \Lambda \mid}$ is substituted in the third line. We obtain the new objective function {${\rm tr} \left( {\bf A}^{\rm H}{\bf A}{\bf A}^{\rm H}{\bf A} \right)$}  in (\ref{Equ:optcode2}) instead of the former function {${\rm tr}\left( \left( {\bf A}^{\rm H}{\bf A} \right)^{-1} \right)$} in (\ref{Equ:optcode}). It is easier to obtain the analytic gradient of the objective function in (\ref{Equ:optcode2}); see the ensuing subsection. Two objective functions in (\ref{Equ:optcode}) and (\ref{Equ:optcode2}) have similar variation trends with respect to $\bf c$, which is demonstrated by numerical examples in Section \ref{Sec:sim}.
\par
As mentioned in the preceding paragraphs, both optimizations in (\ref{Equ:optcode}) and (\ref{Equ:optcode2}) are making the sub-dictionary $\bf A$ approach a semi-unitary matrix {${\bf A}^{\rm H}{\bf A} \rightarrow {\bf I}_{|\Lambda|}$, which means} the columns of $\bf A$ are optimized to be approximately perpendicular to each other.
{In a radar system, target returns which are highly correlated can interfere each other and are difficult to distinguish. Such returns can affect the reconstruction performance. In the context of RSF radar, a column in $\bf A$ represents a normalized echo from a target, and interference between returns of two targets can be characterized by correlation between the columns in $\bf A$, i.e. $|A_i^{\rm H}A_j|$, $i \neq j$. Higher correlation result indicates that more intensive interference exist between the target returns.
Adaptively designed $\bf A$, in which the columns are perpendicular to each other, implies that echoes from different targets are orthogonal, and the interferences among target returns are thus avoided.
The orthogonality contributes to better recovery of targets.}
\par
{In \cite{Gogineni2011} and \cite{Sen2011}, different criterions are applied for adaptive waveforms design to improve sparse recovery performance.
In \cite{Gogineni2011}, the criterion is to maximize the minimum target returns. However, it is not applicable to the RSF radar. We assume that all radar pulses are transmitted at an invariant peak power. Thus, the power of the transmissions is not adjustable and the intensities of the target returns are independent of the transmitted waveforms.
In \cite{Sen2011}, the criterion is based on minimizing an upper bound on $\left\| \mathbf{x}-\widehat{\mathbf{x}} \right\|_{2}^{2}$. However, the upper bound is not tight enough. Decreasing upper bound does not directly imply decrease in the recovery errors.
In this paper, a lower bound on MSE, i.e. CRLB, is chosen as the optimum criterion to design the carrier frequencies. A lower bound defines the best performance that an estimator can obtain. The lower bound equals the MSE if the bound is achievable by some estimators. The achievability of CRLB is discussed in the last paragraph of Section II. Decreasing this bound can result in decrease in MSE.
Note that both \cite{Gogineni2011} and \cite{Sen2011} concern with the power allocation of the waveforms. The methods involved in these two papers are not directly applicable to RSF radar, because the power of the transmissions in the RSF radar is non-adjustable.}
\subsection{Batch-oriented Code Design}{\label{subsec:steepest}}
We consider the situation in which a batch of codes are simultaneously designed; see Fig. \ref{fig:batch}. Since the bandwidth of transmitted waveforms is limited, the code sequence $\bf c$ in RSF radar is optimized by solving the constrained minimization problem
\begin{equation}{\label{Equ:optcodeconstrain}}
\min \limits_{\bf c} {\rm tr} \left( {\bf A}^{\rm H}{\bf A}{\bf A}^{\rm H}{\bf A} \right),\text{ subject to } {\bf 0} \preceq {\bf c} \preceq {\bf 1}.
\end{equation}
where $\bf 0$ and $\bf 1$ denote a vector with all entries 0 and 1, respectively.\par
Our first step is to rewrite (\ref{Equ:optcodeconstrain}) as an unconstrained problem with variable substitutions. Since $f(z) = \arctan(z)$ has a definition domain $z \in (-\infty {{,}}\ \infty)$ and a range $f(z) \in (-\pi/2 {{,}}\ \pi/2)$, we can apply the arctan function for variable substitution to eliminate the domain constraints on the code sequence $\bf c$ in (\ref{Equ:optcodeconstrain}). Relax the constraints {$c_n \in [0,1]$} as {$c_n \in \left(-\delta, {1{+}\delta}\right)$}, where $\delta$ is a small positive constant, $n = 0,1,\dots,N-1$. Introduce a vector ${\bf z} = [z_0,z_1,\dots,z_{N-1}]^{\rm T} \in \mathbb{R}^{N}$, where
\begin{equation}{\label{Equ:zn}}
z_n = \tan \left( \frac{\pi}{{{1}}+2\delta} \left(c_n - \frac{{{1}}}{2}\right) \right),
\end{equation}
or equivalently,
\begin{equation}{\label{Equ:zn1}}
c_n =\frac{{{1}}}{2} + \frac{{{1}}+2\delta}{\pi}\arctan{z_n}.
\end{equation}
Substitute (\ref{Equ:zn1}) into (\ref{Equ:optcodeconstrain}); thus, {the matrix} $\bf A$ becomes a function of $\bf z$ and (\ref{Equ:optcodeconstrain}) is replaced with an unconstrained optimum problem
\begin{equation}{\label{Equ:optcodenoconstrain}}
\min \limits_{\bf z} {\rm tr} \left( {\bf A}^{\rm H}{\bf A}{\bf A}^{\rm H}{\bf A} \right), {\bf z} \in \mathbb{R}^{N}.
\end{equation}
\par
The second step is to calculate the gradient ${\partial {\rm tr}{\left( {\bf A}^{\rm H}{\bf A}{\bf A}^{\rm H}{\bf A} \right)}}/{\partial {\bf z}}$ such that steepest descent method can be applied to solve (\ref{Equ:optcodenoconstrain}). The gradient is
\begin{equation}{\label{Equ:gradient}}
\frac{\partial {\rm tr}{\left( {\bf A}^{\rm H}{\bf A}{\bf A}^{\rm H}{\bf A} \right)}}{\partial {\bf z}} = 4 {\rm Re} \left[ {\rm diag}\left( {\bf D}{\bf A}^{\rm H}{\bf A}{\bf A}^{\rm H} \right) \right],
\end{equation}
where the $n$th row, $i$th column element of ${\bf D} \in \mathbb{C}^{N \times |\Lambda|}$ is $D(n,i) = {{\rm d}A(n,i)}/{{\rm d}z_n}$. ${\rm Re} [\cdot]$ denotes the real part of a complex number. The derivation is presented in the Appendix.
\par
For the steepest descent method, the computational complexity is determined by the required number of iterations and the computational load in each iteration. The convergence speed is discussed with numerical experiments in Subsection \ref{subsec:simsteepest}.
In each iteration, calculating the steepest descent direction with (\ref{Equ:gradient}) and searching for the step size along the direction compose the main computational load.
Calculating the gradient vector (\ref{Equ:gradient}) requires
${3 \mathord{\left/  {\vphantom {3 2}} \right.  \kern-\nulldelimiterspace} 2}N|\Lambda {|^2} + N|\Lambda |$
times complex-valued multiplication and addition.
The locally optimal step size is decided by the values of cost function (\ref{Equ:optcodenoconstrain}) and the calculation requires ${1 \mathord{\left/  {\vphantom {1 2}} \right.  \kern-\nulldelimiterspace} 2}N|\Lambda {|^2} + |\Lambda|$  times complex-values multiplication and addition.
\subsection{Sequential Code Design}\label{subsec:sequential}
In this subsection, we discuss algorithms that design the frequency-modulation codes sequentially. In each design operation, only one code is optimized; see Fig. \ref{fig:sequential}. Suppose $N$ codes ${\bf c} \in \mathbb{R}^{N}$ have been utilized, and measurements ${\bf y} \in \mathbb{C}^{N}$ are available. Our goal is to determinate the next code $c_{N+1}$ with the previous data and the corresponding estimate results.\par
With the measurements $\bf y$, assume that we obtain the estimate of the support set as $\Lambda$ and the corresponding sub-dictionary as ${\bf A} = {\boldsymbol \Phi}_{\Lambda} \in \mathbb{C}^{N \times |\Lambda|}$. When the new code $c_{N+1}$ is applied, a new row ${\bf a}^{\rm H} \in \mathbb{C}^{1 \times |\Lambda|}$ is added to the sub-dictionary $\bf A$. Denote ${\bf A}_{\rm new} = \left[ {\bf A}^{\rm H} \ {\bf a} \right]^{\rm H}$ as the new sub-dictionary; thus, the objective function for sequential code design is
\begin{equation}{\label{Equ:optcode1_seq}}
c_{N+1} = \mathop {\arg \min}  {\rm tr} \left( \left({\bf A}_{\rm new}^{\rm H}{\bf A}_{\rm new} \right)^{-1} \right),
\end{equation}
or approximately,
\begin{equation}{\label{Equ:optcode2_seq}}
c_{N+1} = \mathop {\arg \min} {\rm tr} \left( {\bf A}_{\rm new}^{\rm H}{\bf A}_{\rm new}{\bf A}_{\rm new}^{\rm H}{\bf A}_{\rm new} \right).
\end{equation}
Let ${\bf B} = {\bf A}^{\rm H}{\bf A}$ and ${\bf F} = {\bf B}^{-1}$.  Since
\begin{equation}
\begin{split}
\left({\bf A}_{\rm new}^{\rm H}{\bf A}_{\rm new} \right)^{-1} = \left({\bf B} + {\bf a}{\bf a}^{\rm H} \right)^{-1} = {\bf F} - \frac{{\bf F}{\bf a}{\bf a}^{\rm H}{\bf F}}{1+{\bf a}^{\rm H}{\bf F}{\bf a}},
\end{split}
\end{equation}
the optimum problem (\ref{Equ:optcode1_seq}) can be simplified as
\begin{equation}{\label{Equ:opt1_seq}}
\begin{split}
c_{N+1} &= \arg \min {\rm tr}\left( {\bf F} \right) - \frac{{\rm tr}\left( {\bf F}{\bf a}{\bf a}^{\rm H} {\bf F}\right)}{1+{\bf a}^{\rm H}{\bf F}{\bf a}} \\
&= \arg \max \frac{{\bf a}^{\rm H} {\bf F} ^{2}{\bf a}}{1+{\bf a}^{\rm H}{\bf F}{\bf a}}.
\end{split}
\end{equation}
The approximation (\ref{Equ:optcode2_seq}) can be rewritten as
\begin{equation}{\label{Equ:opt2_seq}}
\begin{split}
c_{N+1} &= \arg \min {\rm tr}\left( {\bf B}^{2}+ {\bf a}{\bf a}^{\rm H}{\bf B}+ {\bf B}{\bf a}{\bf a}^{\rm H}+ {\bf a}{\bf a}^{\rm H}{\bf a}{\bf a}^{\rm H} \right)  \\
&= \arg \min {\rm tr}\left( {\bf B}^{2}\right)+ 2{\bf a}^{\rm H}{\bf B}{\bf a}+ \left({\bf a}^{\rm H}{\bf a}\right)^{2} \\
&= \arg \min {\bf a}^{\rm H}{\bf B}{\bf a}.
\end{split}
\end{equation}
where ${\bf a}^{\rm H}{\bf a} = |\Lambda|$; see (\ref{Equ:echogrid}).\par
Sequential code design with (\ref{Equ:opt1_seq}) or (\ref{Equ:opt2_seq}) is computationally efficient.
Both (\ref{Equ:opt1_seq}) and (\ref{Equ:opt2_seq}) have simple and analytical expressions, and they are single-parameter (i.e., $c_{N+1}$) optimization problems.
Exhaustive search or other numerical methods \cite{Gill1981} can be applied to solve them.
Since there is no matrix-inverse operation involved in (\ref{Equ:opt2_seq}), the approximated objective function in (\ref{Equ:optcode2_seq}) is more computationally efficient than the original function in (\ref{Equ:optcode1_seq}).

\section{Simulation Results}{\label{Sec:sim}}
Numerical results demonstrate the merits of the proposed cognitive scheme. We choose an RSF radar where the central frequency $f_c = 10$ GHz, {the synthetic bandwidth $B = 40$ MHz, pulse duration $T_p = 0.1$ $\mu$S.
We assume that $\Delta f \rightarrow 0$; thus, the codes are continuous parameters. The impacts of the discretization of the codes are discussed in Subsection \ref{subsec:simseq}.
The width of a coarse range bin is $cT_p/2$.
The resolution of HRRP is $c/(2B)$. Then there are $M = (cT_p/2)/(c/2B)=4$ high-range-resolution cells in a coarse range bin.} Unless specifically noted, a coherent processing interval (CPI) consists of $N =20$ radar pulses. We set the number of possible high-resolution rang cells $P = M$, and the number of possible Doppler cells $Q = N$.
The noise are Gaussian white noise and we define the normalized signal to noise ratio as SNR$_i = \mid \gamma_i \mid^2/(N\sigma^2)$ with respect to $i$th target.
Since we focus on applying cognitive idea to RSF radar in this paper, we simply assume that the sparsity level $K$ is known a priori in the SP method. However, in some practical situations, $K$ is unknown. For those cases, we could first estimate the number of targets by combining SP with some well-known order-selection criteria, e.g., the Akaike's information criterion (AIC) \cite{Akaike1974}, the Kullback-Leibler information criterion (KIC) \cite{Cavanaugh1999}, and the minimum description length (MDL) criterion \cite{Rissanen1978}.
In all of the following examples, SP iterates no more than 50 times.
\subsection{Performance of Optimization Criterion} {\label{subsec:simcriterion}}
We discuss whether a code sequence with a lower constrained CRB leads to a smaller MSE. There are $K = 4$ targets with uniform scattering coefficients $\gamma _1 = \gamma _2 = \gamma _3  = \gamma _4 = 1$. The {{range}} and motion parameters of the targets are set such that the parameters of the high-resolution range are $p_1 = 3\Delta p$, $p_2 = 2\Delta p$, $p_3 = 3\Delta p$, $p_4 = 0$, and Doppler parameters are $q_1 = 7\Delta q$, $q_2 = 13 \Delta q$, $q_3 = 14\Delta q$, $q_4 = 15 \Delta q$; see (\ref{Equ:echo}) for the definitions of $p$ and $q$.
The normalized SNR of each target is SNR$ = 20$ dB. We randomly generate 100 code sequences, ${\bf c}^1,{\bf c}^2,\dots,{\bf c}^{100}$. All of the codes are independently and uniformly distributed {between 0 and 1}. For each sequence ${\bf c}^i$, we calculate the normalized, constrained CRB ${\rm LB}({\bf c}^i) = {\rm tr} \left( ({\bf A}^{\rm H}{\bf A})^{-1} \right)$ in (\ref{Equ:optcode}) and estimate the sparse vector ${\bf x}$ with SP in 1000 dependent Monte-Carlo trials. Then, we calculate MSE$({\bf c}^i)$, the MSE of the estimates with code sequence ${\bf c}^i$, and plot MSE$({\bf c}^i)$ versus ${\rm LB}({\bf c}^i)$ in Fig. \ref{fig:msevstr}.
\par
\begin{figure}[!h]
\centering
\includegraphics[width=3in]{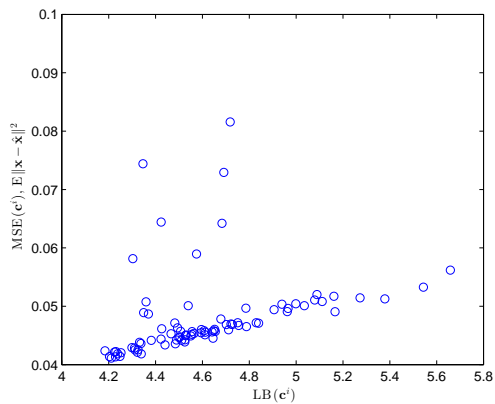}
\caption{MSE of $\bf x$ estimates versus the normalized, constrained CRB ${\rm LB}({\bf c})$. Each blue circle in the figure represents a code sequence ${\bf c}$.}
\label{fig:msevstr}
\end{figure}
As shown in Fig. \ref{fig:msevstr}, a code with lower constrained CRB most likely leads to a lower MSE, so lowering the constrained CRB could be a valid strategy to reduce the recovery error.
{As discussed in Section \ref{Sec:cog}, the matrix $\bf A$ with lower CRB indicates a more informative sensing system, or indicates that interferences among target returns are relieved from the perspective of radar system.}
\subsection{Approximation of the Original Objective Function}
In Subsection \ref{subsec:obj} we approximate the original objective ${\rm tr} \left( ({\bf A}^{\rm H}{\bf A})^{-1} \right)$ in (\ref{Equ:optcode}) with ${\rm tr} \left( {\bf A}^{\rm H}{\bf A}{\bf A}^{\rm H}{\bf A} \right)$ in (\ref{Equ:optcode2}) for efficient computation of the gradient vector. In this subsection, numerical experiments are presented to demonstrate that the approximation is reasonable in the scenario of RSF radar. We randomly generate 2000 code sequences ${\bf c}^{i}$ with an uniform distribution {between 0 and 1}, $i = 1, 2, \dots, 2000$. The target scene is the same as that described in Subsection \ref{subsec:simcriterion}. Then, we compare values of two objective functions ${\rm LB}({\bf c}^i) = {\rm tr} \left( ({\bf A}^{\rm H}{\bf A})^{-1} \right)$ and ${\rm LB}_2({\bf c}^i) = {\rm tr} \left( {\bf A}^{\rm H}{\bf A}{\bf A}^{\rm H}{\bf A} \right)$ with respect to ${\bf c}^i$. The relationship of the two functions is shown in Fig. \ref{fig:trvstr}. We could see that the approximation ${\rm LB}_2({\bf c}^i)$ has a trend similar to the original function ${\rm LB}({\bf c}^i)$. The original objective trends to descend when ${\rm LB}_2({\bf c}^i)$ is reduced. Note that the lower limits of both functions are $K=4$.
\par
\begin{figure}[!h]
\centering
\includegraphics[width=3in]{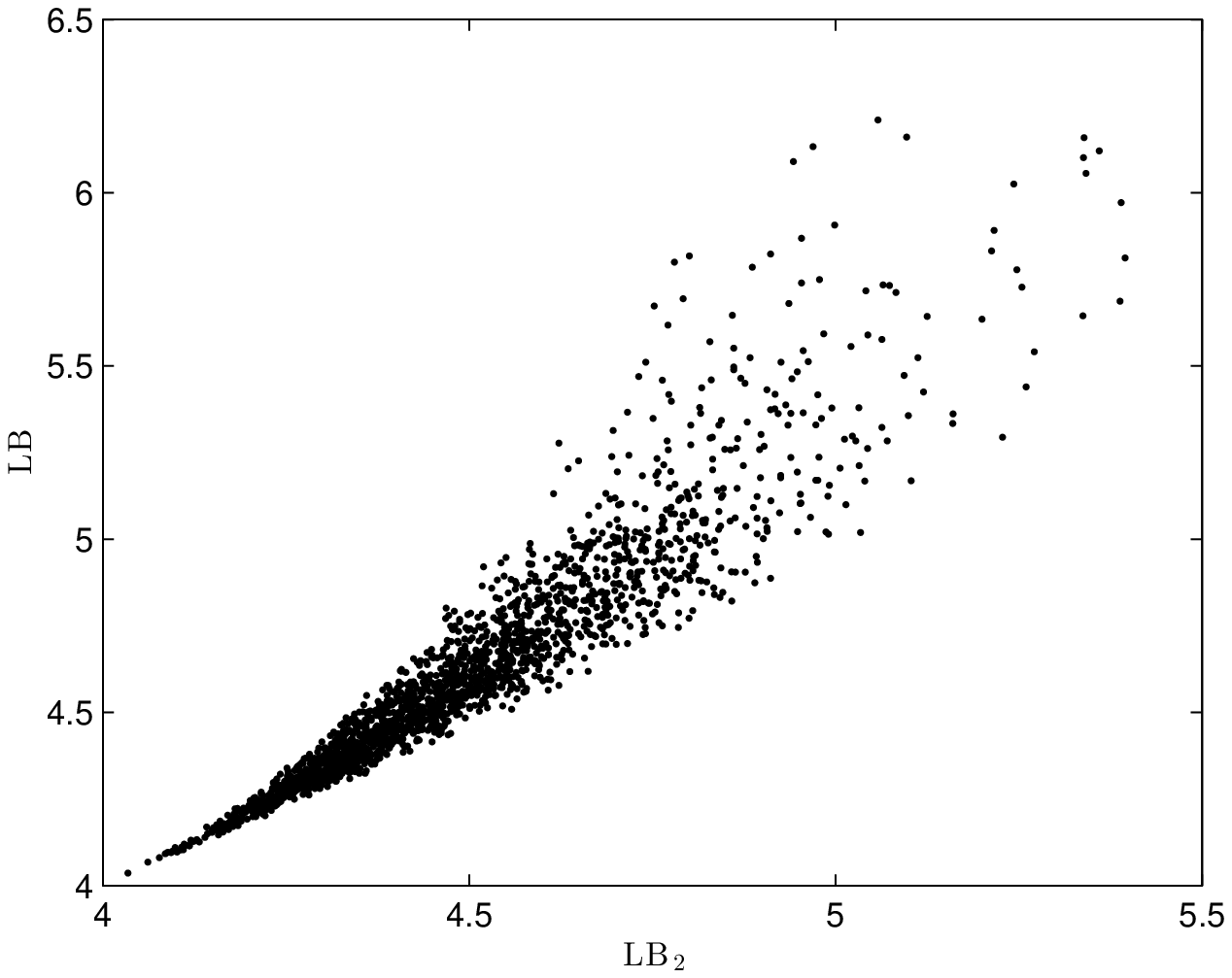}
\caption{${\rm LB}({\bf c}^i) = {\rm tr} \left( ({\bf A}^{\rm H}{\bf A})^{-1} \right)$ versus ${\rm LB}_2({\bf c}^i) = {\rm tr} \left( {\bf A}^{\rm H}{\bf A}{\bf A}^{\rm H}{\bf A} \right)$. The plot contains 2000 black dots, each of which representes a code sequence ${\bf c}^i$.}
\label{fig:trvstr}
\end{figure}
\subsection{Convergence of the Steepest Descent Method} {\label{subsec:simsteepest}}
In this subsection, we discuss the convergence of the proposed code design method in Subsection \ref{subsec:steepest}. The target scheme is the same as that in Subsection \ref{subsec:simcriterion}. The initial code sequences are randomly created, each of which obeys an uniform distribution {between 0 and 1}. Then, these code sequences are optimized under the objective function (\ref{Equ:optcodenoconstrain}) with the steepest descent method, which iterates no more than 100 times. The small constant $\delta$ in (\ref{Equ:zn}) equals 0.1. We calculate the mean of values of the objective function ${\rm LB}_2({\bf c}) = {\rm tr} \left( {\bf A}^{\rm H}{\bf A}{\bf A}^{\rm H}{\bf A} \right)$ versus the iteration counter $l$ in the steepest descent method. The plot are presented in Fig. \ref{fig:trvsl}. The results show that the devised algorithm rapidly reduces values of ${\rm LB}_2({\bf c})$ and closely converges to $K =4$, which is the lower limit of ${\rm tr} \left( {\bf A}^{\rm H}{\bf A}{\bf A}^{\rm H}{\bf A} \right)$.
The time complexity of the algorithm is also tested on a personal computer with Intel$^{\circledR}$ Core$^{\rm TM}$ 2 Duo CPU 3 GHz, 4 GB RAM. The codes are run by MATLAB$^{\circledR}$ 2012. It takes 8.5 seconds to perform 100 Monte-Carlo trials with 100 iterations in each trail.
\par
\begin{figure}[!h]
\centering
\includegraphics[width=3in]{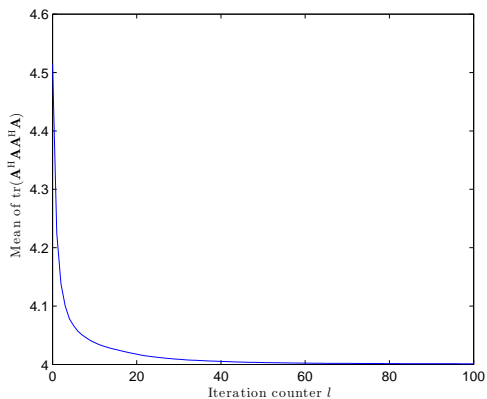}
\caption{The mean of ${\rm tr} \left( {\bf A}^{\rm H}{\bf A}{\bf A}^{\rm H}{\bf A} \right)$ versus the iteration counter $l$ of the steepest descent method. The curve is obtained via averaging the results of 100 Monte-Carlo trials.}
\label{fig:trvsl}
\end{figure}
\par
\subsection{Performance of the Batch-oriented Cognitive Scheme}
In this subsection, simulations are executed to test the reconstruction performance with the cognitive mechanism that implements the steepest descent method in Subsection \ref{subsec:steepest}. We consider two successive CPIs to demonstrate the merits of the cognitive scheme.
In the first CPI, a predefined pseudo-random code sequence ${\bf c}^{(0)}$ is applied, and we obtain the estimate ${\hat{\bf x}}^{(0)}$ and the corresponding support set ${\Lambda}^{(0)} = {\rm supp}\left({\hat{\bf x}}^{(0)}\right)$.
In the second CPI, we simulate the predefined mode and the adaptive mode, respectively, and compare the corresponding results. In the predefined mode, ${\bf c}^{(0)}$ is used again. In the latter mode, optimal code sequences are used, which are calculated via the steepest descend method with ${\Lambda}^{(0)}$.
The target scene is the same as that in Subsection \ref{subsec:simcriterion}. The variance of the Gaussian white noise is varied such that the normalized SNR changes. In the examples, we examine the reconstruction errors and the fractions of exactly recovered support sets.
\par
As shown in Fig. \ref{fig:comparison}, the adaptive mode leads to lower MSEs and higher fractions of exactly recovered support sets than does the predefined mode.
\begin{figure*}[!t]
\centering
\subfloat[MSEs]
{\includegraphics[width=3in]{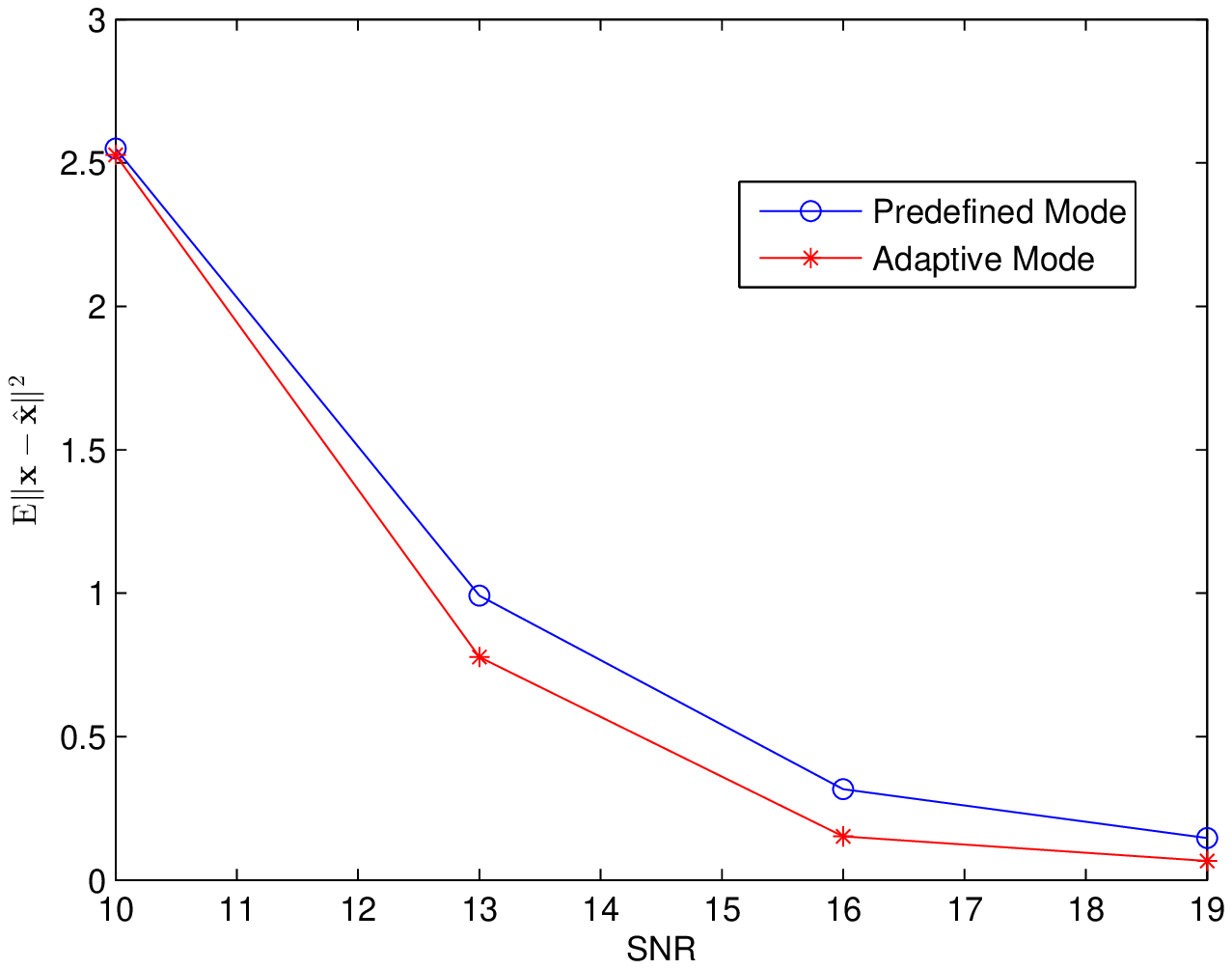}\label{fig:errvssnr}}
\hfil
\subfloat[Fractions of exactly recovered support sets]
{\includegraphics[width=3in]{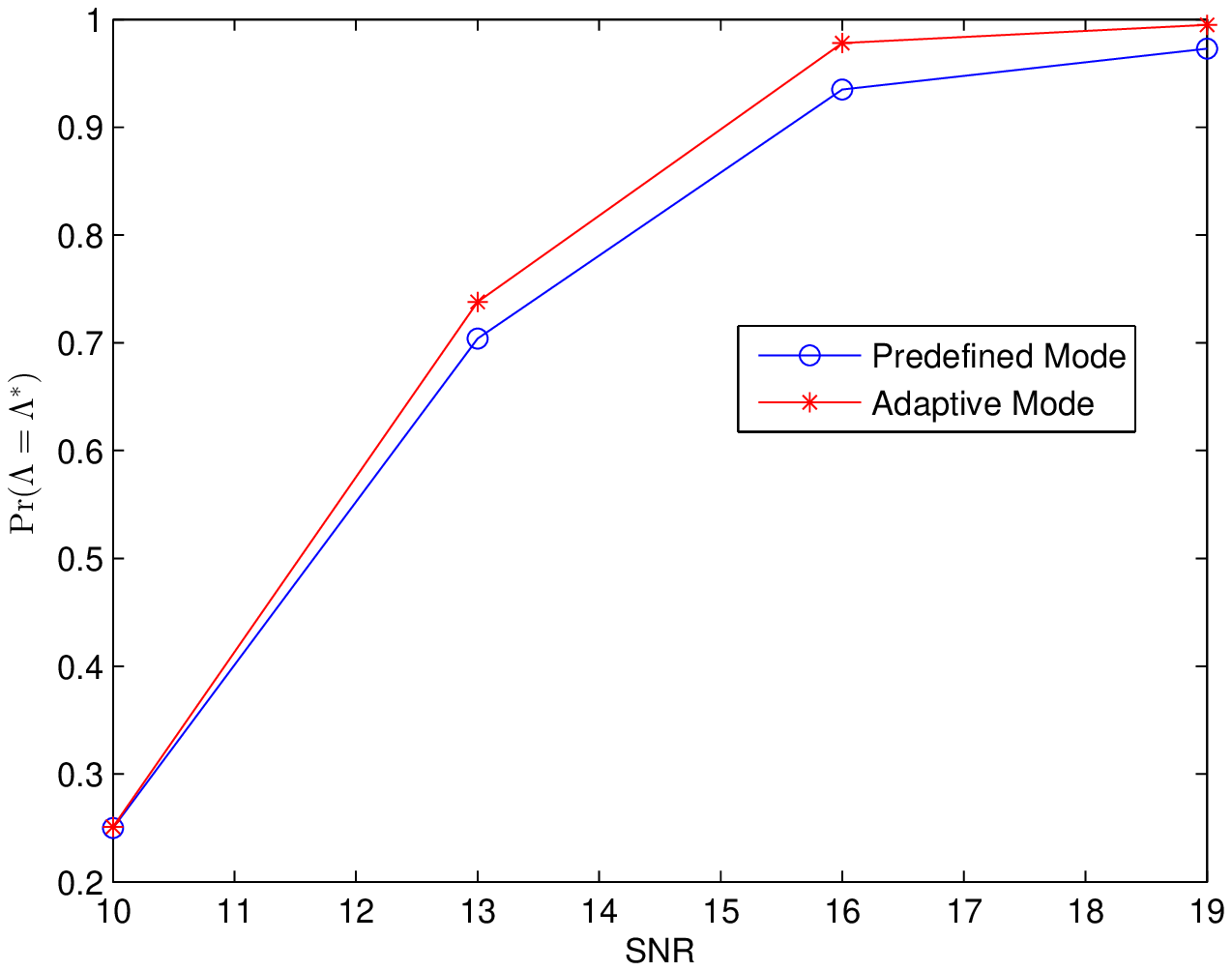}\label{fig:prvssnr}}
\caption{The MSEs and the fractions of exactly recovered support sets with predefined and adaptive code sequences versus SNR. Each circle-asterisk pair in the plots is obtained from 1000 independent Monte-Carlo trials.}
\label{fig:comparison}
\end{figure*}
\subsection{Performance of the Sequential Cognitive Scheme}{\label{subsec:simseq}}
In this subsection, we consider the cognitive mechanism that uses the sequential code design algorithms in Subsection \ref{subsec:sequential}. The target scene is the same as that in Subsection \ref{subsec:simcriterion}. The initial $N=20$ modulation codes ${\bf c}^{(0)}$ are randomly drawn from an uniform distribution over $[0,1]$. We test the adaptive modes and the random mode. In the adaptive modes, we sequentially designed the ensuing 20 codes with the methods in Subsection \ref{subsec:sequential}, while in the random mode the ensuing codes are drawn randomly as comparison. Once a new code is determined, the recovery errors are calculated. As discussed in Subsection \ref{subsec:sequential}, two methods are applicable for the adaptive modes. We denote the code design processes with (\ref{Equ:opt1_seq}) and (\ref{Equ:opt2_seq}) as 'Adaptive Mode 1' and 'Adaptive Mode 2', respectively.\par
The reconstruction errors and fractions of exactly recovered support sets are depicted in Fig. \ref{fig:seq}, where the noise variances are set as $\sigma^2 = 0$ dB or $\sigma^2 = 5$ dB. Two adaptive modes have similar recovery errors, which are much smaller than those of the random mode. Fractions of exactly recovered support sets of both adaptive modes are close to each other, which are much higher than the random mode.
{The time complexity of these three modes are tested on a personal computer with Intel$^{\circledR}$ Core$^{\rm TM}$ 2 Duo CPU 3 GHz, 4 GB RAM. The codes are run by MATLAB$^{\circledR}$ 2012. For 1000 Monte-Carlo trials, the consumed time of 'Adaptive Mode 1' and 'Adaptive Mode 2' are 37.0 and 34.5 seconds, respectively, which are very close to 33.7 seconds, the counterpart of 'Random Mode'. This indicates that the code optimum processes consume minor computational efforts.}
Note that Adaptive Mode 2 has lower computation load than Adaptive Mode 1, and it may be more suitable for some real-time applications.\par
\begin{figure*}[!h]
\centering
\subfloat[MSEs, $\sigma^2 = 0$ dB]
{\includegraphics[width=3in]{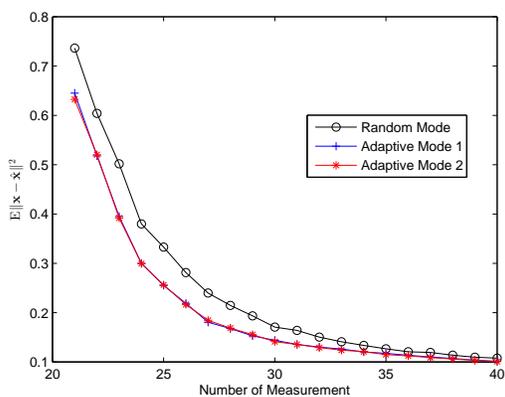}\label{fig:mse0dB}}
\hfil
\subfloat[Fractions of exactly recovered support sets, $\sigma^2 = 0$ dB]
{\includegraphics[width=3in]{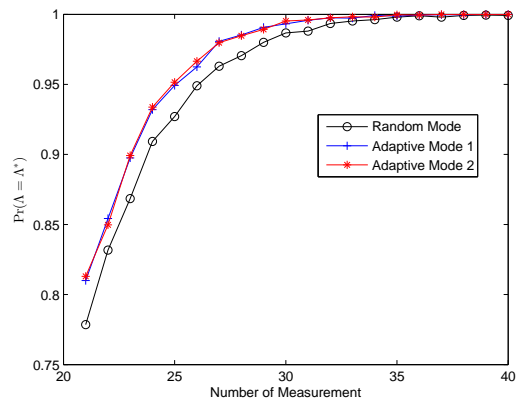}\label{fig:pr0dB}}
\hfil
\subfloat[MSEs, $\sigma^2 = 5$ dB]
{\includegraphics[width=3in]{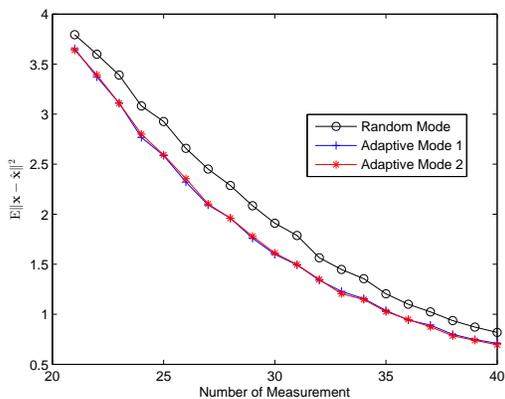}\label{fig:msem5dB}}
\hfil
\subfloat[Fractions of exactly recovered support sets, $\sigma^2 = 5$ dB]
{\includegraphics[width=3in]{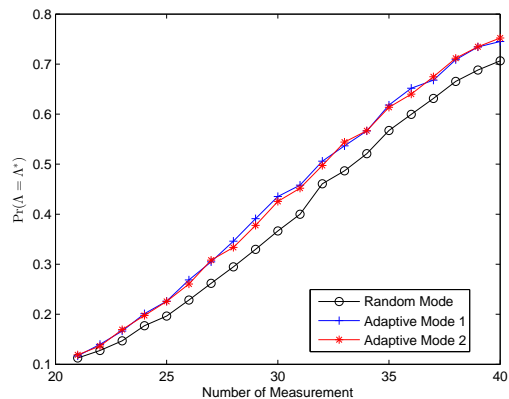}\label{fig:prm5dB}}
\caption{The MSEs and the fractions of exactly recovered support sets with random and adaptive code sequences versus number of measurements. In (a) and (b), the noise variance $\sigma^2 = 0$ dB, and in (c) and (d) $\sigma^2 = 5$ dB. In 'Random Mode', the codes are randomly generated. In 'Adaptive Mode 1' and 'Adaptive Mode 2', the codes are sequentially optimized by (\ref{Equ:opt1_seq}) and (\ref{Equ:opt2_seq}), respectively. Each circle or asterisk in the plots is obtained from 1000 independent Monte-Carlo trials.}
\label{fig:seq}
\end{figure*}
Then we consider the influence of the frequency step size $\Delta f$. The performances of the three modes are evaluated with respect to different $\Delta f$.
There are 4 targets, of which the ranges and velocities are randomly generated. The scattering intensities of the targets are all one. The initial $N=20$ codes are randomly generated. The MSEs are calculated after the ensuing 5 pulses are transmitted. The noise variance $\sigma ^2 = 0$ dB. The results are shown in Fig. \ref{fig:msevsdeltaf}, and the adaptive modes outperform the nonadaptive mode for all tested $\Delta f$.
\par
\begin{figure}[!h]
\centering
\includegraphics[width=3in]{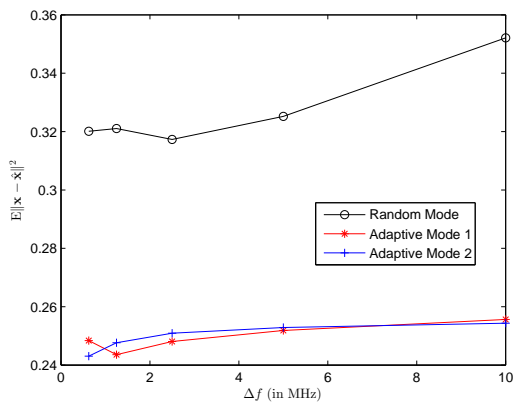}
{\caption{The MSEs with random and adaptive code sequences versus the frequency step size $\Delta f$. The results are obtained from 20000 independent Monte-Carlo trials.}\label{fig:msevsdeltaf}}
\end{figure}
We also discuss impacts of the number of targets $K$ on the performance of the cognitive mechanism. The target scene and the initial $N=20$ codes are randomly generated. The scattering intensities of the targets are all one. We calculate the recovery errors after the ensuing 5 codes are designed. The noise variance $\sigma ^2 = -5$ dB. The results are shown in Fig. \ref{fig:msevsk}, and we can see that the adaptive modes produce lower recovery errors for all $K$ that appear in Fig. \ref{fig:msevsk}.
\begin{figure}[!h]
\centering
\includegraphics[width=3in]{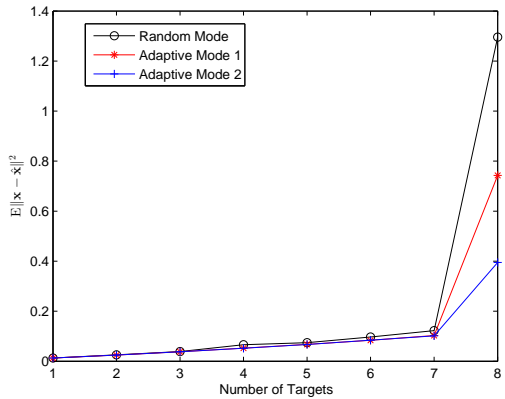}
\caption{The MSEs with random and adaptive code sequences versus the number of targets $K$. The results are obtained from 4000 independent Monte-Carlo trials.}
\label{fig:msevsk}
\end{figure}
\section{Conclusion}{\label{Sec:conclusion}}
In this paper, we develop a novel notion of cognitive random stepped frequency radar that employs compressed sensing algorithm to reconstruct the target scene. We propose a new criterion and several effective methods to adaptively optimize the carrier frequencies of the radar pulses. With the information about the observed targets, the proposed cognitive mechanism significantly eliminates the interference between target returns and improves the sensing performance of radar. Numerical simulations are performed to demonstrate that the devised cognitive system reduces the recovery errors of the targets.\par
\appendix
We derive the gradient of $f({\bf z}) = {\rm tr}{({\bf A}^{\rm H}{\bf A}{\bf A}^{\rm H}{\bf A})}$ with respect to ${\bf z}$.
Denote $a_{ml}$ and $b_{ml}$ as the $m$th row, $l$th column element of ${\bf A}$ and ${\bf B}={\bf A}^{\rm H}{\bf A} \in \mathbb{C}^{\mid\Lambda\mid \times \mid\Lambda\mid}$, respectively. We have that
\begin{equation}{\label{Equ:dcdz}}
\frac{{\rm d} c_n}{{\rm d} z_n} = \frac{{{1}}+2\delta}{\pi(z_n^2+1)},
\end{equation}
\begin{equation}{\label{Equ:dadc}}
\frac{\partial a_{ml}}{\partial c_n} = \left\{
\begin{array}{l}
0,m \neq n,\\
a_{nl}(jp_l+jq_lm{{B}}/f_c),m=n,
\end{array}
\right.
\end{equation}
\begin{equation}{\label{Equ:dbdc}}
\begin{split}
\frac{\partial b_{ml}}{\partial c_n} &= \frac{\partial \sum \limits_i a_{im}^{*}a_{il}}{\partial c_n} = \frac{\partial a_{nm}^{*}a_{nl}}{\partial c_n}\\
&= \left(\frac{\partial a_{nm}}{\partial c_n}\right)^{*}a_{nl} +  a_{nm}^{*} \frac{\partial a_{nl}}{\partial c_n},
\end{split}
\end{equation}
\begin{equation}{\label{Equ:dtrdc}}
\frac{\partial {\rm tr}\left({{\bf B}^{\rm H}{\bf B}}\right)}{\partial c_n} = \sum \limits_{m,l} \frac{\partial b_{ml}^{*}b_{ml}}{\partial c_n} = 2{\rm Re} \left[ \sum \limits_{m,l}  \frac{b_{ml}^{*}\partial b_{ml}}{\partial c_n}  \right].
\end{equation}
{With the substitution of} (\ref{Equ:dcdz} - \ref{Equ:dtrdc}), we get
\begin{equation}{\label{Equ:dtrdz}}
\begin{split}
\frac{\partial f({\bf z})}{\partial z_n} &= \frac{\partial {\rm tr}\left({{\bf B}^{\rm H}{\bf B}}\right)}{\partial c_n} \cdot \frac{{\rm d} c_n}{{\rm d} z_n}\\
&= 2{\rm Re} \left[ \sum \limits_{m,l} d_{nm}b_{ml}a_{nl}^{*} + a_{nm}b_{ml}d_{nl}^{*} \right],
\end{split}
\end{equation}
where
\begin{equation}{\label{Equ:dadz}}
d_{nl} {=} \frac{{\rm d}a_{nl}}{{\rm d}z_n} {=} \frac{{\rm d}a_{nl}}{{\rm d}c_n} \frac{{\rm d}c_{n}}{{\rm d}z_n} {=} a_{nl}(jp_l{+}\frac{jq_lm{{B}}}{f_c}) \frac{{{1}}+2\delta}{\pi(z_n^2{+}1)}
\end{equation}
{denotes the $n$th row, $l$th column element of matrix $\bf D$}.
Rewrite (\ref{Equ:dtrdz}) in matrix form,
\begin{equation}{\label{Equ:dtrdzmatrix}}
\begin{split}
\frac{\partial f({\bf z})}{\partial {\bf z}} &{=} 2{\rm Re} \left[ {\rm diag} \left( {\bf DBA}^{\rm H} {+} {\bf ABD}^{\rm H} \right) \right] \\
&{=} 4{\rm Re} \left[ {\rm diag} \left( {\bf DBA}^{\rm H} \right) \right].
\end{split}
\end{equation}
\section*{Acknowledgment}
This work was supported in part by the National Natural Science
Foundation of China (No. 60901057 and No. 61201356) and in part by the National Basic
Research Program of China (973 Program, No. 2010CB731901).\par
The authors would like to thank Prof. Arye Nehorai for his insightful comments on various versions of this manuscript, anonymous reviewers for their important suggestions, and also Mr. Yuanxin Li who contributed some suggestions on writing Subsection \ref{subsec:obj}.

\ifCLASSOPTIONcaptionsoff
  \newpage
\fi



\bibliography{CognizantRSF}
%

%
\begin{biography}[{\includegraphics[width=1in,height=1.25in,clip,keepaspectratio]{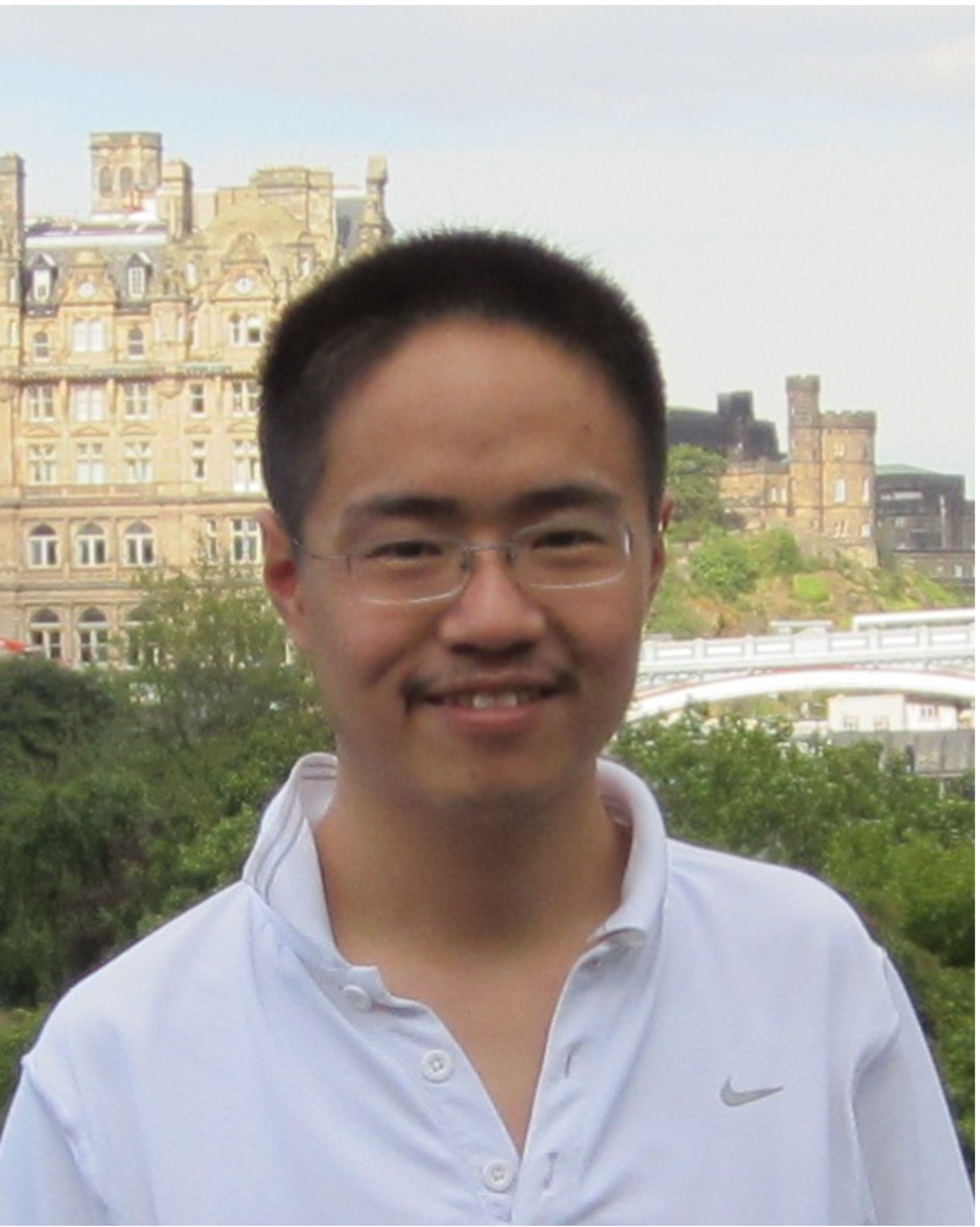}}]{Tianyao Huang}
received the B.S. degree in electronic and information engineering from Harbin Institute of Technology, Harbin, Heilongjiang, China, in 2009. \par
He is currently pursuing his Ph.D. degree in electronic engineering at Tsinghua University, Beijing, China. From 2012 to 2013, he was a visiting student with Dr. Wei Dai at Imperial College London, London, UK. His research interests are radar signal processing, compressed sensing and electronic system design.
\end{biography}
\begin{biography}[{\includegraphics[width=1in,height=1.25in,clip,keepaspectratio]{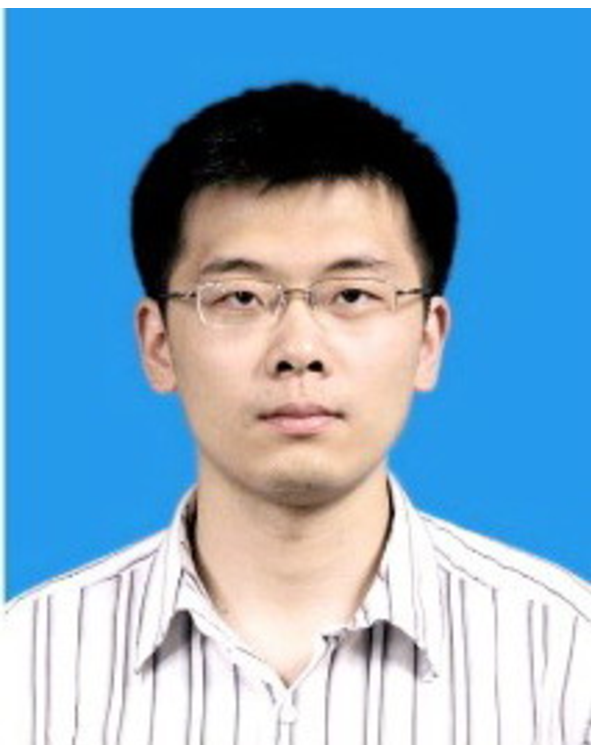}}]{Yimin Liu}
(M'12) received the B.S. and Ph.D degrees (both with honors) in electronics engineering from the Tsinghua University, Beijing, China, in 2004 and 2009, respectively.
\par
From 2004, he was with the Intelligence Sensing Lab. (ISL), Department of Electronic Engineering, Tsinghua University. He is currently an assistant professor with Tsinghua, where his field of activity is research in new concept radar and other microwave sensing technology. His current research interests include radar theory, statistic signal processing, compressive sensing and their applications in radar, spectrum sensing and intelligent transportation systems.
\end{biography}
\begin{biography}[{\includegraphics[width=1in,height=1.25in,clip,keepaspectratio]{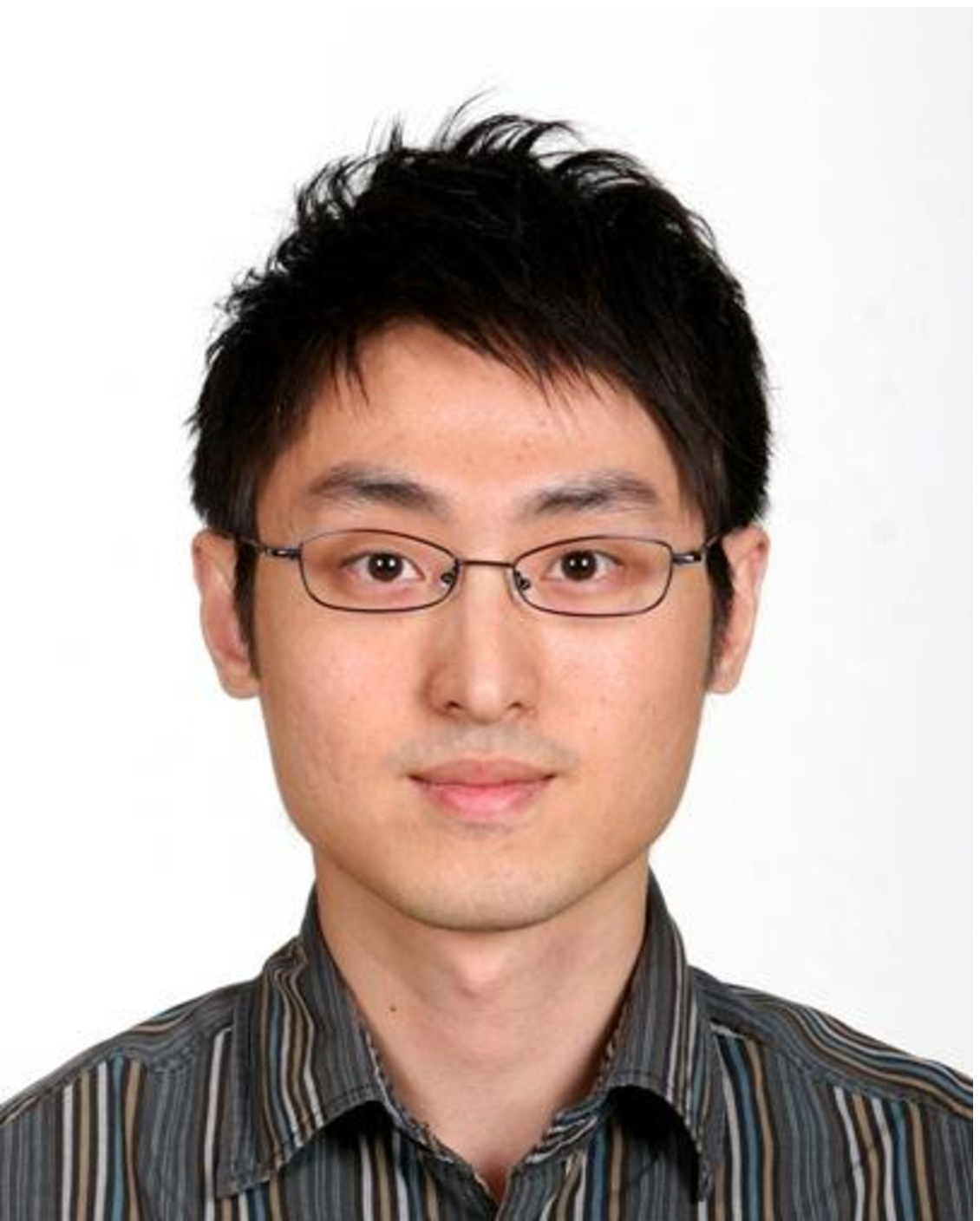}}]
{Huadong Meng}
(S'01-M'04) received the B.Eng. degree and Ph.D. degree in electronic engineering, both from Tsinghua University, Beijing, China, in 1999 and 2004 respectively. \par
In 2004, he joined the Faculty of Tsinghua University, where he is currently an Associate Professor in the Department of Electronic Engineering. He is a member of the Technical Committee of the 2013 IET International Radar Conference. His current research interests include statistical signal processing, target tracking, sparse signal processing, cognitive radar, and spectrum sensing in cognitive radio networks.
\end{biography}
\begin{biography}[{\includegraphics[width=1in,height=1.25in,clip,keepaspectratio]{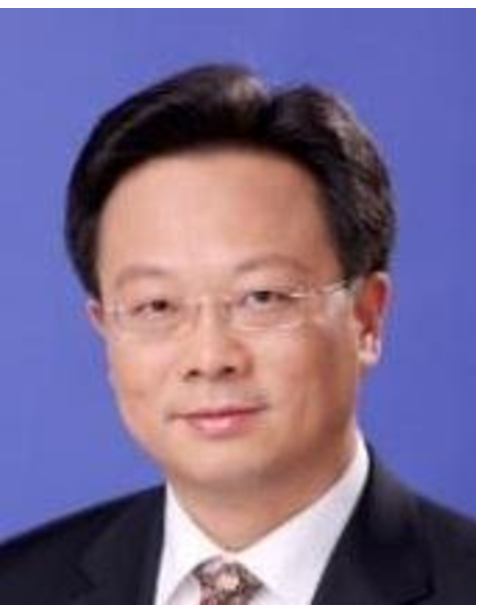}}]{Xiqin Wang}
received his B.S. and Ph.D. degrees in Electronic Engineering from Tsinghua University, Beijing, China, in 1991 and 1996 respectively. He was a visiting scholar and assistant research engineer of PATH/ITS, UC  Berkeley from 2000 to 2003. \par
He is currently a professor with the Department of Electronic Engineering of Tsinghua University, and also chaired the department from 2006-2012. His current research interests include radar and communications signal processing, image processing, compressed sensing and cognitive signal processing, and electronic system design. He is also interested in structure of knowledge and curricula reforming in electronic engineering.
\end{biography}
%
%
%




\end{document}